\documentclass[floatfix,
reprint,
superscriptaddress,
 amsmath,amssymb,
 aps,
pra,
]{revtex4-2}
\DeclareUnicodeCharacter{025B}{\ensuremath{\varepsilon}}
\usepackage{gensymb}
\usepackage{graphicx}
\usepackage{dcolumn}
\usepackage{bm}
\usepackage{physunits}
\usepackage{xcolor}
\usepackage{amsmath}

\begin{document}

\preprint{APS/123-QED}

\title{
Photo-induced switching of magnetisation in the epsilon-near-zero regime}


\author{Héloïse Damas}
\email{heloise.damas@ru.nl}
\affiliation{
HFML-FELIX Laboratory, Radboud University, Toernooiveld 7, 6525 ED Nijmegen, The Netherlands 
}

\author{Carl S. Davies}
\email{carl.davies@ru.nl}
\affiliation{
HFML-FELIX Laboratory, Radboud University, Toernooiveld 7, 6525 ED Nijmegen, The Netherlands 
}

\author{Petr M. Vetoshko}
\affiliation{Russian Quantum Center, Skolkovo Innovation Center, Bolshoi Blv.,
30, bl.1, Moscow, 121205, Russia}

\author{Vladimir I. Belotelov}
\affiliation{Russian Quantum Center, Skolkovo Innovation Center, Bolshoi Blv.,
30, bl.1, Moscow, 121205, Russia}
\affiliation{Faculty of Physics, Lomonosov Moscow State University, Leninskie Gory, Moscow 119992, Russia}

\author{Andrzej Stupakiewicz}
\affiliation{Faculty of Physics, University of Bialytsok, 1L Ciolkowskiego, 15-245 Bialystok, Poland}

\author{Andrei Kirilyuk}
\email{andrei.kirilyuk@ru.nl}
\affiliation{
HFML-FELIX Laboratory, Radboud University, Toernooiveld 7, 6525 ED Nijmegen, The Netherlands 
}

\begin{abstract}
The possibility of controlling spins using ultrashort light and strain pulses has triggered intense discussions about the mechanisms responsible for magnetic re-ordering. All-optical magnetisation switching can be achieved through ultrafast heat-driven demagnetisation or transient modifications of magnetic anisotropy. During the phononic switching of magnetic dielectrics, however, mid-infrared optical excitations can modify the crystal environment via both the thermal quenching of anisotropy and the generation of strain respectively, with the relative distinction between these thermal and non-thermal processes remaining an open question. Here, we examine the effect of mid-infrared pulses tuned to the frequency of optical phonon resonances on the labyrinthine domain structure of a cobalt-doped yttrium iron garnet film. We find that the labyrinthine domains are transformed into stable parallel stripes, and quantitative micromagnetic calculations demonstrate this stems predominantly from a partial quenching of the anisotropy. Contrary to conventional wisdom, however, we find that this heat-facilitated process of magnetisation switching is spectrally strongest not at the maximum of absorbed optical energy but rather at the epsilon-near-zero points. Our results reveal that the epsilon-near-zero condition provides an alternative pathway for laser-driven control of magnetisation, even when the underlying mechanism is primarily thermal. 
\end{abstract}

\maketitle


The manipulation of magnetisation using light represents a topic of considerable interest from both technological and fundamental perspectives. Optical control over magnetisation offers the prospect of ultrafast, low-dissipation control of magnetic order, an essential capability for high-speed data storage technologies, while also facilitating complex magnetisation dynamics, yielding access to non-equilibrium trajectories and phases that are otherwise unattainable \cite{beaurepaire1996ultrafast, radu2011transient}. 

Since its first demonstration in 2007 \cite{stanciu2007all}, all-optical magnetisation switching has been reported in a variety of material systems and achieved through different mechanisms. These include single-shot helicity-independent switching in ferrimagnets \cite{mentink2012ultrafast, davies2022helicity}, helicity-dependent switching in thin films with large spin-orbit coupling \cite{lambert2014all}, and photomagnetic switching driven by the resonant excitation of specific electronic transitions in iron-garnets \cite{stupakiewicz2017ultrafast, zalewski2024ultrafast}. Such diverse processes can be straightforwardly classified as thermal and non-thermal, depending on the impact of heat on the magnetisation during the switching process. For example, exchange-driven switching in ferrimagnetic alloys is driven entirely by ultrafast heating \cite{mentink2012ultrafast, davies2020pathways}, involving a transient destruction and subsequent recovery of the net magnetisation, whereas photomagnetic switching in iron garnets relies instead on electronic transitions that are activated at resonance \cite{stupakiewicz2017ultrafast, zalewski2024ultrafast}, resulting in the precessional reorientation of the magnetisation vector. 

Non-thermal switching mechanisms have predominantly relied on the transient modification of anisotropy thereby altering the potential energy profile governing magnetisation dynamics. In this context, nonlinear phononics has recently emerged as a powerful approach for modifying crystal structure and fields through the resonance of infrared-active optical phonon modes \cite{forst2011nonlinear,subedi2014theory, mankowsky2016non}. In magnetic systems, the strain and magnetoelastic fields arising fron such lattice excitations can exert a direct influence on magnetisation \cite{nova2017effective, disa2020polarizing}.

In 2021, Stupakiewicz \emph{et al.} \cite{stupakiewicz2021ultrafast} demonstrated that resonant excitation of longitudinal optical phonon modes in iron garnets can induce permanent magnetisation reversal, referred to as \textit{phononic switching}. In that study, switched magnetic domains emerged at the periphery of the illuminated area, where the head load is minimal, suggesting a non-thermal origin. This specific spatial distribution was somewhat explained by the generation of strain and the associated spatially-varying magnetoelastic fields and torque, an interpretation supported qualitatively by micromagnetic simulations \cite{stupakiewicz2021ultrafast, gidding2023dynamic}. On the one hand, as the experiments were conducted at equilibrium several seconds after the end of the magnetisation dynamics, the precise spatio-temporal extent of the underlying spin-lattice interactions remains unresolved. On the other hand, it is also well-known that the thermal load delivered by optical pulses in magnetic dielectrics can reduce magnetocrystalline anisotropy more strongly than magnetisation \cite{davies2019anomalously}. Under appropriate conditions, this can drive large-amplitude precession of magnetisation and even reversal. The possible contribution of thermally induced change of anisotropy during the \textit{phononic switching} of magnetisation thus remains an open question.

In this paper, we examine the influence of mid-infrared excitations on the labyrinthine domain structure of cobalt-doped yttrium iron garnet (Co:YIG) films. Our magneto-optical microscopy-based measurements reveal that a single optical excitation at a particular wavelength induces a transformation of the magnetic domains from the initial labyrinth-like configuration to a stable arrangement of parallel stripes. Moreover, upon applying a modest magnetic field along certain directions during the irradiation, the labyrinthine magnetic domains instead transform into stable magnetic bubbles. Using quantitive micromagnetic simulations, incorporating both thermal and non-thermal interactions on an equal footing, we are able to fully explain all aspects of the observed process predominantly in terms of a thermally-induced transient reduction of the magnetocrystalline anisotropy. However, spectrally-resolved measurements reveal that the thermally-induced magnetic structure transformation scales not with the absorptance and pump energy profile of Co:YIG but rather with the inverse of its dielectric function. In other words, the strongest thermal effect is obtained under the epsilon-near-zero (ENZ) condition, not at the maximum of absorptance and pump energy where the strongest thermal effects are expected. These findings therefore reveal the importance of the ENZ condition in enhancing light-matter interaction, even for thermally-driven processes, and establish it as offering an alternative pathway for all-optical magnetisation switching.

\begin{figure}[h!]
  \centering
  \includegraphics[width=0.95\linewidth]{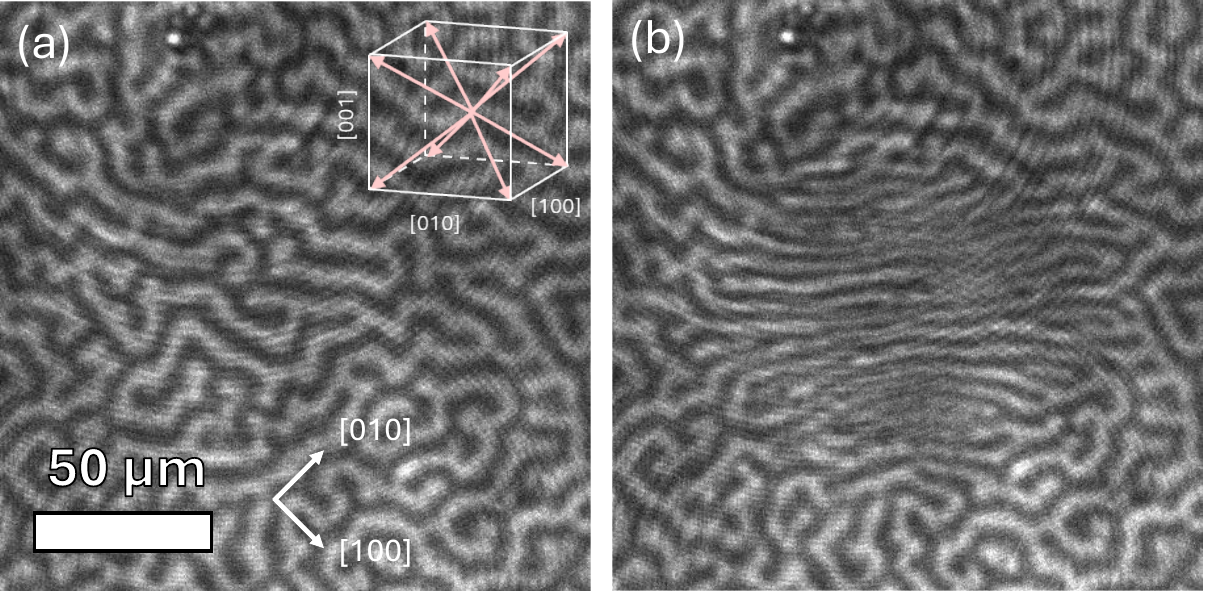}
  \caption{Faraday images of the magnetic domain structure probed with wavelength $\lambda = 1064$ nm (a) before and (b) after illumination with a single burst of pulses (wavelength $\lambda =$ 13 $\mu$m). (a - inset) Schematic of the cubic crystal structure with the four equivalent [111] magnetocrystalline easy axes.}
  \label{fig:1}
  \end{figure}


\noindent \textbf{\large{Results}} \\
We study here a cobalt-doped film of yttrium-iron-garnet (Co:YIG) of thickness 13 $\mu$m grown atop a substrate of gadolinium gallium garnet (see Methods). The insertion of Co results in a strong magnetocrystalline anisotropy, consisting of cubic and uniaxial components that favour an orientation of magnetisation close to the body diagonal of the cubic cell \cite{marysko1994anisotropy,maryvsko1995cubic,tekielak1997magnetic}, as shown in the inset of Fig.~\ref{fig:1}(a). For Co:YIG in the absence of any external magnetic fields, the interplay between shape and magnetocrystalline anisotropies gives rise to a labyrinthine domain structure, as shown in Fig.~\ref{fig:1}(a) with a domain width of $w^{labyrinth} = 5\,\mu$m. This type of magnetic domain structure is quite common in garnets \cite{hubert2008magnetic}. Our experiments are performed at the free-electron laser facility FELIX in Nijmegen, the Netherlands \cite{oepts1995free}, which delivers bursts of $\approx200$ transform-limited pulses coming at 25 MHz with a central wavelength (photon energy) from 7 $\mu m$ (177 meV) to 22 $\mu m$ (56 meV) (see Methods and Supp.~1). A single burst of pulses was selected using a fast mechanical shutter and focused onto the Co:YIG sample at normal incidence using an off-axis parabolic mirror. Meanwhile, the magnetic domain structure was imaged by Faraday microscopy using 6-ns-long pulses of wavelength $\lambda = 1064$ nm for illumination, as delivered by an Nd:YAG laser  (see Supp.~2 for a schematic of the pump-probe setup). \\

Prior to irradiation, the magnetic domain structure is initialised by applying and subsequently removing a saturating in-plane magnetic field. Upon exposing the sample to a single 8-$\mu$s-long macropulse, with a central wavelength of 13 $\mu$m, we observe the creation of stripe domains oriented parallel to each other (see Fig.~\ref{fig:1}(b)). These stripe domains are entirely stable. This experiment was repeated for different initialised magnetisation orientations along the equivalent [111] axes, as set by applying saturating in-plane magnetic fields. The orientation of the parallel stripes appears to follow the in-plane component of the magnetisation; in other words, the stripes align with the in-plane projection of the closest easy axes. Note that the width of the stripes is $w^{stripes} \approx 3-4 \, \mu$m, which is smaller than the initial labyrinth domain width. Also note that the switching is independent of the relative angle between the pump polarisation and the easy axes.

\begin{figure}
  \centering
  \includegraphics[width=0.93\linewidth]{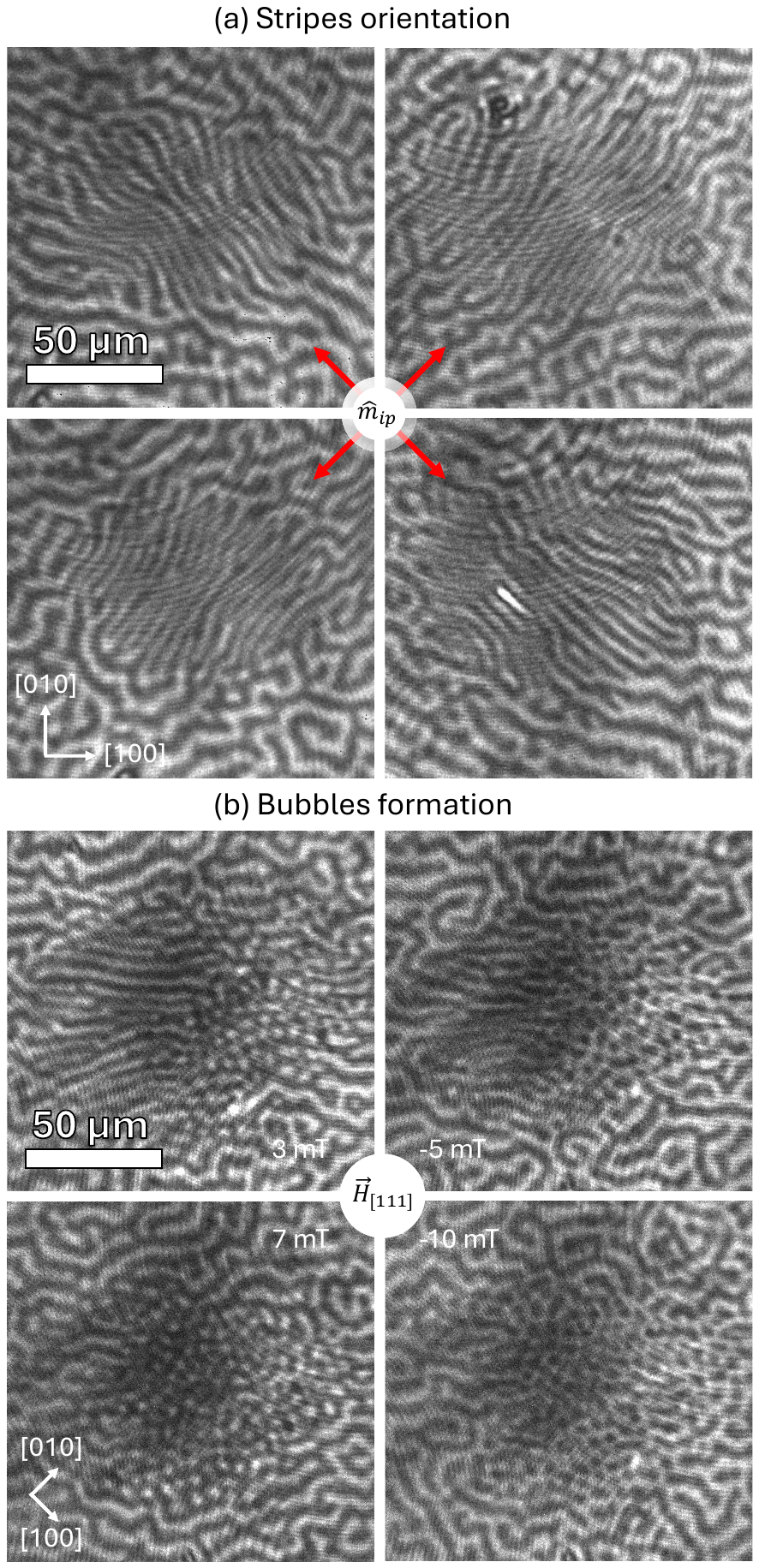}
  \caption{(a) Faraday images obtained after illumination with the pump, for various initial magnetisation orientations described by the vector $\hat{m}_{ip}$. The orientation of the parallel stripes follows that of the initial in-plane projection of the magnetisation. (b) Faraday images obtained after illumination with the pump, with an external magnetic field applied along an easy axis.}
  \label{fig:switching}
\end{figure}

Illuminating the sample while applying a magnetic field along an easy axis results in the formation of stable magnetic bubbles (see Fig.~\ref{fig:switching}(b)). For a modest magnetic field, the magnetic bubbles coexist with the stripe domains, while slightly larger magnetic fields lead to the dominance of the magnetic bubbles. The contrast of the magnetic bubbles is reversed for an opposite field. Surprisingly, the bubbles remain stable even when the magnetic field is turned off.\\


\begin{figure}[ht]
\centering
\includegraphics[width=0.93\linewidth]{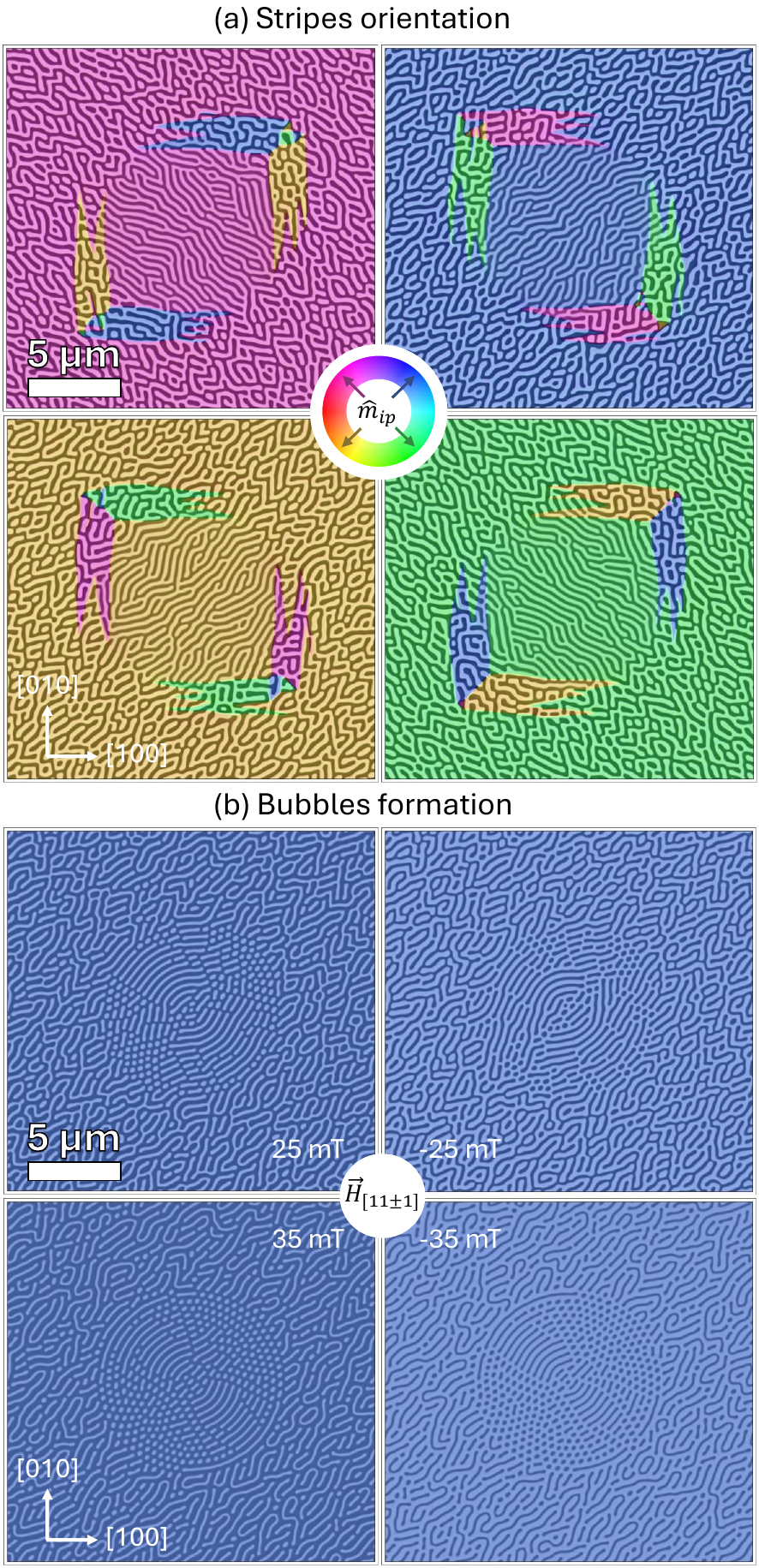}
\caption{(a) Simulated magnetic state 50 ns after the strain pulse with a 75\% change in magnetocrystalline anisotropy, for different initial magnetisation orientations. (b) Simulated magnetic state 100 ns after the stain pulse with a 70 \% change in magnetocrystalline anisotropy and while an external magnetic field is applied along [111] or [11-1].}
\label{fig:simulations}
\end{figure}

\noindent \textbf{\large{Discussion}} \\
Previous experiments have shown that a static strain, induced by placing a loaded sphere on garnet films, can modify the labyrinthine domain structure into parallel stripes \cite{van1978scratch}. To therefore attempt to explain the observed transformation of magnetic domains, we performed complementary micromagnetic modelling using Mumax3 \cite{vansteenkiste2014design}. We take into account both exchange and magnetostatic interactions, and adopt reasonable sample parameters (see Methods). However, our micromagnetic simulations, which include only the effect of a transient strain and are quantitatively comparable to experiment, do not reproduce stripe formation (see Supp.~3). Instead, they only reveal switching of in-plane domains at the periphery of the irradiated area, resembling the results presented by Stupakiewicz \textit{et al.} \cite{stupakiewicz2021ultrafast}.


To proceed further, we recall that the experiments were performed with an 8-$\mu$s-long burst of pulses at a 25 MHz repetition rate, which could likely incur accumulative heating. Moreover, the iron-garnet film studied here exhibits stronger uniaxial anisotropy and higher saturation magnetisation compared to the iron-garnet studied in Ref. \cite{stupakiewicz2021ultrafast}. To overcome the energy barrier for reversal, the necessary strain would thus need to be larger, which similarly introduces stronger heating effects. 
We therefore next consider the possibility that thermal reduction of magnetocrystalline anisotropy could drive the observed domain transformation \cite{davies2019anomalously}. In Co:YIG, we expect both the cubic and uniaxial magnetocrystalline anisotropy constants to decrease with increasing temperature \cite{pardavi1985temperature, vertesy2002temperature, vertesy2003temperature, beaulieu2018temperature, panin2025exploring}.

Therefore, micromagnetic simulations were repeated taking into account a transient reduction of the anisotropy constants. The simulations show that reducing the cubic term alone elongates the labyrinth domains, whereas reducing the uniaxial term decreases the out-of-plane component of the magnetisation (see Supp.~4). Reducing both the cubic and uniaxial anisotropy terms together results in the formation of parallel stripes (see Supp.~4 and Supp.~5). Decreasing the magnetisation at saturation increases the domains width (see Supp.~5) while the domains width is decreased in our experiments. Further simulations show that combining a transient reduction in magnetocrystalline anisotropy with transient strain promotes the formation of parallel stripes (see Supp.~6). Fig.~\ref{fig:simulations}(a) shows the simulated magnetic state 50 ns after a transient strain and a 75 \% reduction in magnetocrystalline anisotropy, where parallel stripes form at the centre of the simulated volume, aligned with the in-plane projection of the initial magnetisation.

Going further, we next consider the scenario including the effect of a persistently-applied external magnetic field. When applied with a projection along an easy axis, our simulations reveal the formation of magnetic bubbles. Their contrast switches with inversion of the out-of-plane component of the field (see Fig.~\ref{fig:simulations}(b)), again in complete agreement with our experimental observations.

Taken altogether, the micromagnetic simulations presented in Fig.~\ref{fig:simulations} therefore support the hypothesis that the transformation of the labyrinth domain into parallel stripes and bubbles, observed experimentally, is driven predominantly by a transient reduction in the magnetocrystalline anisotropy, which we attribute to thermal effects. The hypothesis of thermal effects is also supported by the $\mu$s-time-scale on which the magnetisation dynamics takes place (see Supp.~S7).



\begin{figure}
  \centering
  \includegraphics[width=\linewidth]{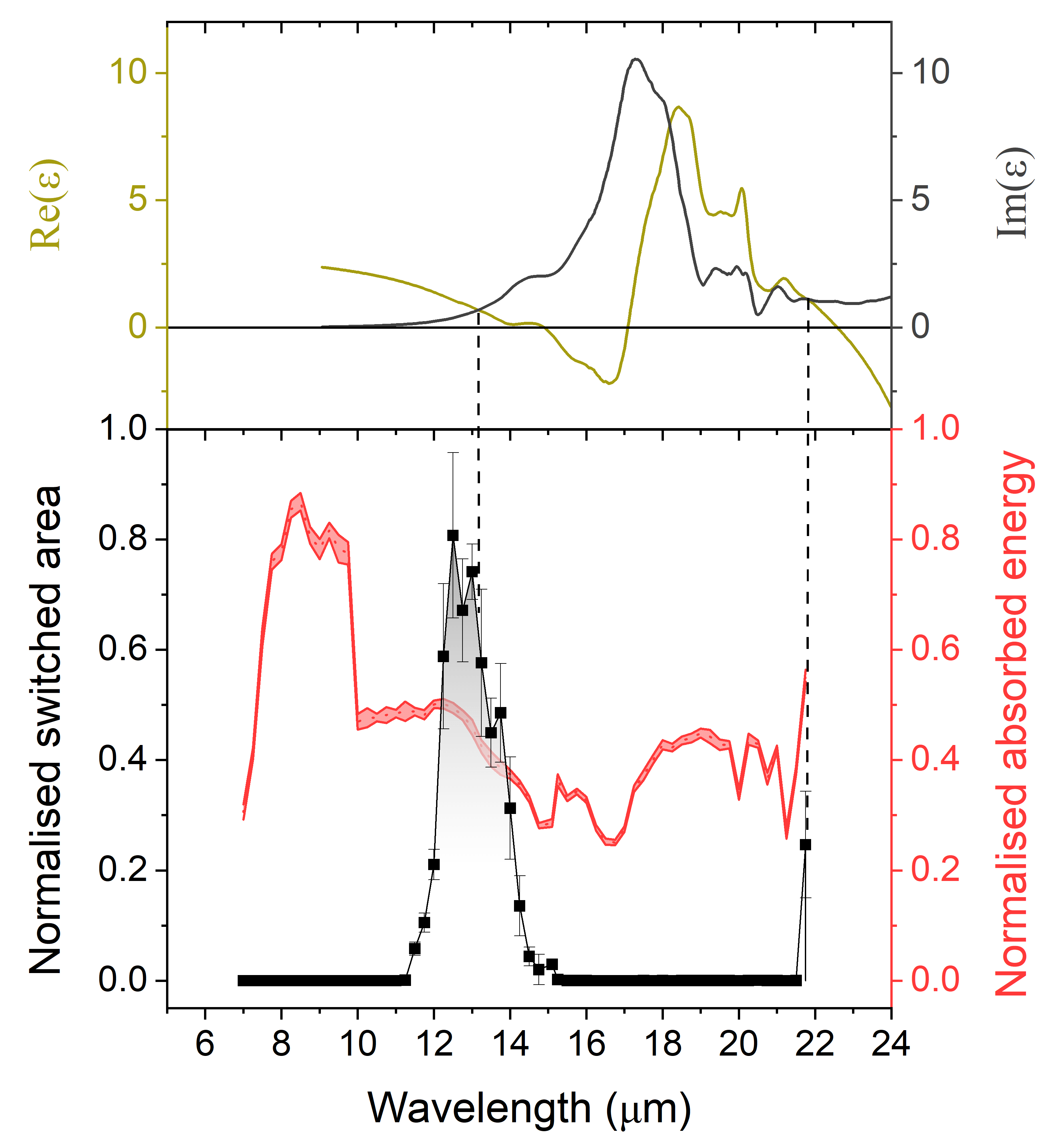}
  \caption{(top panel) Real and imaginary parts of the dielectric function of Co:YIG. (bottom panel) Spectral dependence of the normalised switched area and absorbed optical energy.}
  \label{fig:switching-ENZ}
\end{figure}

Thus far, we have only considered the effect of pumping Co:YIG at a wavelength of 13 $\mu$m. To explore its spectral dependence, the experiment was repeated across a range of pump wavelengths. The results, summarised in the bottom panel of Fig.~\ref{fig:switching-ENZ}, show the normalised switched area, defined as the normalised area of the magnetic stripes. We clearly observe the emergence of a rather narrow maxima about 13 $\mu$m and the beginning emergence of a second peak at 21 $\mu$m. At all other wavelengths, for the same amount of optical energy, the magnetisation is unaffected. To interpret these results, we overlay the absorbed optical energy defined as the incident optical energy multiplied by the Co:YIG absorptance spectrum measured by Fourier transform infrared spectroscopy (FTIR). This reveals a striking observation -- although the micromagnetic simulations indicate that the magnetisation switching is driven by heat-induced changes in magnetocrystalline anisotropy, the strongest effect occurs where the absorbed optical energy is rather modest. Conversely, at the wavelength of the transverse optical phonon mode ($\lambda$ = 17 $\mu$m), where the absorption coefficient reaches its maximum, no switching is observed. Increasing the incident energy at this wavelength only irreversibly damages the crystal once a certain threshold is exceeded (see Supp.~S8).

A direct comparison with the dielectric function  measured by ellipsometry on a similar sample \cite{stupakiewicz2021ultrafast} (top panel of Fig.~\ref{fig:switching-ENZ}) reveals that the switching observed at 13 $\mu$m and 21 $\mu$m correlates with conditions where both the real and imaginary parts of the dielectric function approach zero - known as the epsilon-near-zero (ENZ) points. In the ENZ regime, light exhibits a range of counter-intuitive behaviours \cite{kinsey2019near, reshef2019nonlinear, wu2021epsilon}, including slow light propagation \cite{silveirinha2007theory, niu2018epsilon}, electric field enhancement \cite{campione2013electric}, harmonic generation \cite{vincenti2011singularity}, strong optical nonlinearities \cite{alam2016large, ciattoni2016enhanced, fomra2024nonlinear}, and enhancement of thermo-optical effects \cite{wu2024thermo}. 
Of particular relevance to our results is the ability of the optical electric field to become spatially static across the thickness of the material while remaining temporally dynamic \cite{mahmoud2014wave,xu2019reflection}. Since the time-averaged absorbed power density scales with the product of the absorption coefficient and the square of optical electric field \cite{yariv2007photonics, mayerhofer2021wave}, we infer that at $\lambda$ = 13 $\mu$m the ENZ condition produces a more uniform absorption profile throughout the crystal, distributing the thermal load over a larger volume and thus reducing the risk of damage. In contrast, damage is more likely when pumping near the TO phonon mode resonance, where the highly-absorbing nature of the TO phonon confines the optical absorption to a depth of roughly 500 nm.  
 
The spectral dependence of the magnetisation switching obtained in our study is consistent with observations on a similar Co:YIG film \cite{stupakiewicz2021ultrafast} and antiferromagnetic nickel oxide \cite{stremoukhov2022phononic}. The role of the ENZ regime in switching phenomena was highlighted in the ferroelectric switching of barium titanate \cite{kwaaitaal2024epsilon}, further underscoring the growing recognition of ENZ effects in enabling macroscopic order switching \cite{davies2024epsilon}.

The final issue we address concerns why two nominally similar Co:YIG films display either thermally driven (this study) or strain-driven responses (\cite{stupakiewicz2021ultrafast}). Quantitative micromagnetic simulations of a single domain Co:YIG, comparable to that studied by Stupakiewicz \textit{et al.} \cite{stupakiewicz2021ultrafast}, show that a transient reduction in magnetocrystalline anisotropy leaves the final domain configuration essentially unchanged (see Supp.~S9). Thus, switching driven solely by reduced magnetocrystalline anisotropy would not be discernible in the final magnetic state. However, our simulations also indicate that the same effect can drive the transformation of labyrinthine domains to stripes. In contrast, strain and the associated magnetoelastic fields are expected to act primarily in-plane. Due to the 3 degrees miscut in the sample studied in Ref. \cite{stupakiewicz2021ultrafast}, the influence of magnetoelastic fields on magnetisation can be resolved via Faraday microscopy. In our study, only the out-of-plane component is accessible, and resolving strain effects would require time-resolved experiments combined with detection schemes sensitive to in-plane magnetisation. We therefore propose that both thermal and non-thermal mechanisms operate simultaneously in both samples with possibly different intensities. However, because of differences in magnetic properties and growth conditions, only the strain-driven effect is observable in the Co:YIG of Stupakiewicz \textit{et al.}, whereas only the magnetocrystalline anisotropy reduction is detectable in the presented study.

To summarise, the labyrinthine magnetic domain structure of Co:YIG film is found to change when the material is illuminated with mid-infrared light. The transformation from a labyrinth pattern to parallel stripes or bubbles in the presence of an applied magnetic field can be attributed to a thermal reduction in magnetocrystalline anisotropy as supported by micromagnetic simulations. The wavelength dependence of the magnetisation switching shows that the ENZ condition exerts a stronger influence than the combined effects of material absorptance and pump energy, even when the switching mechanism is thermally driven. These findings highlight the enhanced light–matter interaction characteristic of the ENZ regime. Future experiments employing single micropulses will help eliminate heat accumulation and yield a more complete understanding of the microscopic processes governing magnetisation dynamics.\\

\noindent \textbf{\large{Methods}}\\
\noindent \textbf{Sample}\\
The experiments presented in this paper were carried out on a Y$_{1.94}$Ca$_{1.22}$Fe$_{3.52}$Ge$_{1.31}$Co$_{0.14}$O$_{12}$ (Co:YIG) film of thickness 13 $\mu$m grown by liquid phase epitaxy on a (001)-oriented substrate of gadolinium gallium garnet.

\noindent \textbf{Experimental methodology}\\
To explore the effect of an infrared excitation on the magnetisation, we employed narrow-band mid-infrared pulses from the \textit{Free Electron Lasers for Infrared eXperiments} (FELIX) facility in Nijmegen, the Netherlands \cite{oepts1995free}. This light source delivers bursts (``macropulses'') of optical pulses with a tunable central wavelength between 3 $\mu$m and 1500 $\mu$m. In the present experiments, the excitation wavelength was tuned in the range from 7 $\mu$m to 22 $\mu$m and the measurements were performed at room temperature in air. The 8-$\mu$s-long macropulses are generated at a repetition rate of 10 Hz and contain approximately 200 picosecond-long ``micropulses'' coming at a repetition rate of 25 MHz (see Supp.~1 for a schematic of the time structure of macro- and micro-pulses). For these experiments, a single macropulse was selected using a fast mechanical shutter and focused onto the Co:YIG sample at normal incidence using an off-axis parabolic mirror. Meanwhile, the magnetic domain structure was imaged by Faraday microscopy using 6-ns-long pulses of wavelength $\lambda = 1064$ nm for illumination, as delivered by an Nd:YAG laser  (see Supp.~2 for a schematic of the pump-probe setup). 
\noindent \textbf{Micromagnetic simulations}\\
Micromagnetic simulations were performed using Mumax3 \cite{vansteenkiste2014design}. The total modeled sample volume of size 20 $\mu$m $\times$ 20 $\mu$m $\times$ 1 $\mathrm{\mu}$m was discretised with cells of size 9.77 nm $\times$ 9.77 nm $\times$ 500 nm. The cell size (shorter than the YIG exchange length \cite{sakharov2020spin}) ensures that the exchange energy is accounted for in the plane of the simulated sample. With demagnetisation included, the sample parameters are partially taken from the literature, such that the exchange strength is $A_{exc}=3.7 \times 10^{-12} J/m^3 $ \cite{klingler2014measurements}, the cubic and uniaxial anisotropy constants are respectively $K_c =- 1.5 \times 10^4 \, J/m^3$ and $K_u=5 \times 10^3 J/m^3$ (approximately ten times the values found in the literature \cite{stupakiewicz2017ultrafast}) and the magnetoelastic coupling constants are $b_1 = 3.48 \times 10^5 J/m^3$ and $b_2= 6.96 \times 10^5 J/m^3$ \cite{smith1963magnetostriction}. The saturation magnetisation was set to $M_S=250$ kA/m, and the Gilbert damping was set to $\alpha = 1.2$ (approximately six times the value found in the literature \cite{stupakiewicz2017ultrafast}).\\

\noindent \textbf{\large{Acknowledgments}}\\
We thank the technical staff at FELIX for providing technical support. This publication is part of the project NL-ECO: Netherlands Initiative for Energy-Efficient Computing (with project number NWA. 1389.20.140) of the NWA research programme Research Along Routes by Consortia which is financed by the Dutch Research Council (NWO). C.S.D. acknowledges support from the European Research Council ERC Grant Agreement No. 101115234 (HandShake), and A.K. acknowledges support from the European Research Council ERC Grant Agreement No. 101141740 (INTERPHON).

\bibliography{apssamp}

\end{document}